\documentclass[preprint,tightenlines,showpacs,amsmath,amssymb]{revtex4}
\usepackage{graphicx,bm,amssymb}

\begin{document}

\title{The long journey from the giant-monopole 
       resonance to the nuclear-matter incompressibility}
\author{J. Piekarewicz}
\email{jorgep@csit.fsu.edu}
\affiliation{Department of Physics,
             Florida State University, 
	     Tallahassee, FL 32306}
\date{\today}
\begin{abstract}
Differences in the density dependence of the symmetry energy predicted
by nonrelativistic and relativistic models are suggested, at least in
part, as the culprit for the discrepancy in the values of the
compression modulus of symmetric nuclear matter extracted from the
energy of the giant monopole resonance in ${}^{208}$Pb.  ``Best-fit''
relativistic models, with stiffer symmetry energies than Skyrme
interactions, consistently predict higher compression moduli than
nonrelativistic approaches.  Relativistic models with compression
moduli in the physically acceptable range of $K\!=200\!-\!300$~MeV are
used to compute the distribution of isoscalar monopole strength in
${}^{208}$Pb.  When the symmetry energy is artificially softened in
one of these models, in an attempt to simulate the symmetry energy of
Skyrme interactions, a lower value for the compression modulus is
indeed obtained.  It is concluded that the proposed measurement of the
neutron skin in ${}^{208}$Pb, aimed at constraining the density
dependence of the symmetry energy and recently correlated to the
structure of neutron stars, will also become instrumental in the
determination of the compression modulus of nuclear matter.
\end{abstract}
\smallskip
\maketitle

The compression modulus of symmetric nuclear matter is a fundamental
property of the equation of state.  While some of the existent claims
in the literature may be overstated---indeed, there is little evidence 
in support of a correlation between the compression modulus and the
physics of neutron stars~\cite{Mu96}---the compression modulus impacts
on a diverse set of phenomena ranging from nuclear structure to
supernova explosions. In particular, the compression modulus controls
the energetics around the nuclear-matter saturation point. This is
because the first derivative of the energy-per-nucleon with respect to
the density ({\it i.e.,} the pressure) vanishes at saturation, so
the dynamics of small density fluctuations around the equilibrium
position becomes solely determined by the compression modulus.

To date, most efforts devoted to the study of the compression modulus
have relied on the excitation of the isoscalar giant-monopole
resonance (GMR). While the first set of measurements of the GMR date
back to the late seventies and early eighties~\cite{Yo77,Yo81}, a
recently improved $\alpha$-scattering experiment finds the position of
the giant monopole resonance in ${}^{208}$Pb at $E_{\rm
GMR}\!=\!14.17\!\pm\!0.28$~MeV~\cite{Yo99}.  While the experimental
story on the GMR in ${}^{208}$Pb seems to be coming to an end, the
theoretical picture remains unclear. On the one hand nonrelativistic
calculations that reproduce the distribution of isoscalar-monopole
strength using Hartree-Fock plus random-phase approximation (RPA)
approaches with state-of-the-art Skyrme~\cite{Co92,Ha97} and
Gogny~\cite{Bl95} interactions, predict a nuclear compression modulus
in the range of $K\!=\!210\!-\!220$~MeV~. On the other hand,
relativistic models that succeed in reproducing a large body of
observables, including the excitation energy of the GMR, predict a
larger value for the nuclear incompressibility
($K\!\simeq\!275$~MeV)~\cite{La97,Vr00}. It is the aim of this paper
to elucidate the origin of this apparent discrepancy. It is proposed
that this discrepancy, at least in part, is due to the density
dependence of the symmetry energy; a poorly known quantity that
affects physics ranging from the neutron radius of heavy nuclei to the
structure of neutron stars~\cite{Ho01}. It should be noted that while
knowledge of the symmetry energy is at present incomplete, the
proposed measurement of the neutron radius of ${}^{208}$Pb at the
Jefferson Laboratory~\cite{PREX} should provide stringent constraints
on this fundamental component of the equation of state.

In this paper we follow closely the philosophy of Blaizot and
collaborators who advocate a purely microscopic approach for the
extraction of the compression modulus of nuclear matter from the
energy of the giant-monopole resonance~\cite{Bl95,Bl99}. While the
merit of macroscopic (semi-empirical) formulas for obtaining
qualitative information on the compression modulus is
unquestionable~\cite{Ch97,St98}, the field has attained a level of
maturity that demands stricter standards: it is now expected that
microscopic models predict simultaneously the compression modulus 
of nuclear matter as well as the distribution of isoscalar monopole 
strength. Moreover, theoretical studies based solely on macroscopic
approaches have been proven inadequate~\cite{Pe91,Sh93}.

The starting point for the calculations is an interacting Lagrangian 
density of the following form:
\begin{eqnarray}
{\mathcal L}_{\rm int} = \bar{\psi}\left[
    g_{\rm s}\phi-\left(g_{\rm v}V_{\mu}+\frac{g_{\rho}}{2}
    {\mbox{\boldmath$\tau$}} \cdot {\bf b_{\mu}}+
    \frac{e}{2}(1+\tau_{3})A_{\mu}\right)\gamma^\mu\right]\psi
      -  \frac{\kappa}{3!}(g_{\rm s}\phi)^{3}
      -  \frac{\lambda}{4!}(g_{\rm s}\phi)^{4}\;.
 \label{Lint}
\end{eqnarray}
This Lagrangian includes an isodoublet nucleon field ($\psi$)
interacting via the exchange of scalar ($\phi$) and vector ($V^{\mu}$,
${\bf b}^{\mu}$, and $A^{\mu}$) fields. It also incorporates
scalar-meson self-interactions ($\kappa$ and $\lambda$) that are
instrumental in reducing the unreasonably large value of the
compression modulus predicted in the original (linear) Walecka
model~\cite{Wa74,Se86}. The Lagrangian density depends on five unknown
coupling constants that may be determined from a fit to ground-state
observables. Four of these constants ($g_{\rm s}$, $g_{\rm v}$,
$\kappa$, and $\lambda$) are sensitive to isoscalar observables so
they are determined from a fit to symmetric nuclear matter. The four
nuclear bulk properties selected for the fit are as follows: {\it i)}
the saturation density, {\it ii)} the binding energy per nucleon at
saturation, {\it iii)} the nucleon effective mass at saturation, and
{\it iv)} the compression modulus (see Table~\ref{table1}).  It is
noteworthy, yet little known, that the above four coupling constants
can be determined algebraically and uniquely from these four empirical
quantities~\cite{Gl96}. It is also possible for the various meson
masses to enter as undetermined parameters. However, here the standard
procedure of fixing the masses of the $\omega$ and $\rho$ mesons at
their physical value is adopted; that is, $m_{\rm v}\!=\!783$~MeV and
$m_{\rho}\!=\!763$~MeV. As infinite nuclear matter is only sensitive
to the ratio $g_{\rm s}^{2}/m_{\rm s}^{2}$, the mass of the
$\sigma$-meson must be determined from finite-nuclei properties; 
the $\sigma$-meson mass has been adjusted to reproduce the
experimental root-mean-square (rms) charge radius of ${}^{208}$Pb
($r_{\rm ch}\!=\!5.50\!\pm\!0.01$~fm.)

The symmetry energy of nuclear matter is a poorly known quantity
with an uncontrolled density dependence in nonrelativistic models
(for a recent discussion of the symmetry energy in Skyrme models 
see Refs.~\cite{Br00} and~\cite{Oy02}). In contrast, the symmetry 
energy displays a weak model dependence in relativistic approaches. 
It is given by the following simple form: 
\begin{equation}
  S(k_{\rm F}) = \frac{k_{F}^{2}}{6E_{F}^{*}}
                        + \frac{g_{\rho}^{2}}{12\pi^{2}}
                          \frac{k_{F}^{3}}{m_{\rho}^{2}} \;,
 \label{SymmE}
\end{equation}
where $E_{F}^{*}\!=\!\sqrt{k_{\rm F}^{2}+M^{*2}}$. The symmetry
energy, together with its density dependence, is constrained in
relativistic approaches because the only ``free'' parameter in
Eq.~(\ref{SymmE}) is the $NN\rho$ coupling constant. As the effective
nucleon mass $M^{*}$ has been fixed in symmetric nuclear matter
(and spin-orbit phenomenology demands a value in the range of
$M^{*}/M\!=\!0.6\!-\!0.7$) reproducing the empirical value of the
symmetry energy at saturation ($J\!\simeq\!37$~MeV) constrains the
$NN\rho$ coupling constant to a relatively small range.  Note that
relativistically, the density dependence of the symmetry energy can
also be modified through the inclusion of isoscalar-isovector
couplings terms~\cite{Ho01}, density-dependent coupling
constants~\cite{Ty99}, and isovector-scalar mesons~\cite{Li02}. For
simplicity, however, none of these contributions will be considered
here.  In reality, the symmetry energy at saturation is not well
constrained experimentally. Rather, it is an average of the symmetry
energy near saturation density and the surface symmetry energy that is
constrained by the binding energy of nuclei. Thus, a prescription
first outlined in Ref.~\cite{Ho01} is adopted here: the value of the
$NN\rho$ coupling constant is adjusted, unless otherwise noted, so
that the symmetry energy at $k_F\!=\!1.15$~fm$^{-1}$ 
({\it i.e.,} $\rho\!=\!0.10$~fm$^{-3}$) be equal to $26$~MeV (see
Table~\ref{table1}).

The nuclear observables used as input for the determination of the
model parameters are listed in Table~\ref{table1}. In all cases the
saturation density, binding-energy-per-nucleon, and rms charge radius
in ${}^{208}$Pb have been fixed at their empirical values.  Thus, the
only discriminating factors among the three ``families'' are the
effective nucleon mass and the symmetry energy.  While best-fit
relativistic models suggest values for the symmetry energy and its
slope at saturation density satisfying $J\geq 35$~MeV and $L\geq
100$~MeV, respectively~\cite{Ch97}, family~C is defined with an
artificially small value for $J$ (and correspondingly for $L$) in a
``poor-man's'' attempt at simulating nonrelativistic Skyrme
forces~\cite{Oy02}. That nonrelativistic Skyrme models have a softer
symmetry energy is revealed by the behavior of one of the most
sensitive probes of the density dependence of symmetry energy: the
neutron skin of ${}^{208}$Pb. Indeed, the neutron skin of ${}^{208}$Pb
is predicted to be equal to $R_{n}\!-\!R_{p}\!=\!0.16$~fm for the
recent SkX parametrization and falls below $0.22$~fm for all eighteen
Skyrme parameter sets considered in Ref.~\cite{Br00}. In contrast,
best-fit relativistic models consistently predict larger values. For
example, the NL3 model of Ref.~\cite{La97}, the TM1 model of Sugahara
and Toki~\cite{Su94}, and the NLC model of Serot and
Walecka~\cite{Se97}, predict $R_{n}\!-\!R_{p}\!=\!0.28, 0.27,$ and
$0.26$~fm, respectively (see also Table~\ref{table2}).

\begin{figure}[h]
 \includegraphics[width=1.0\linewidth]{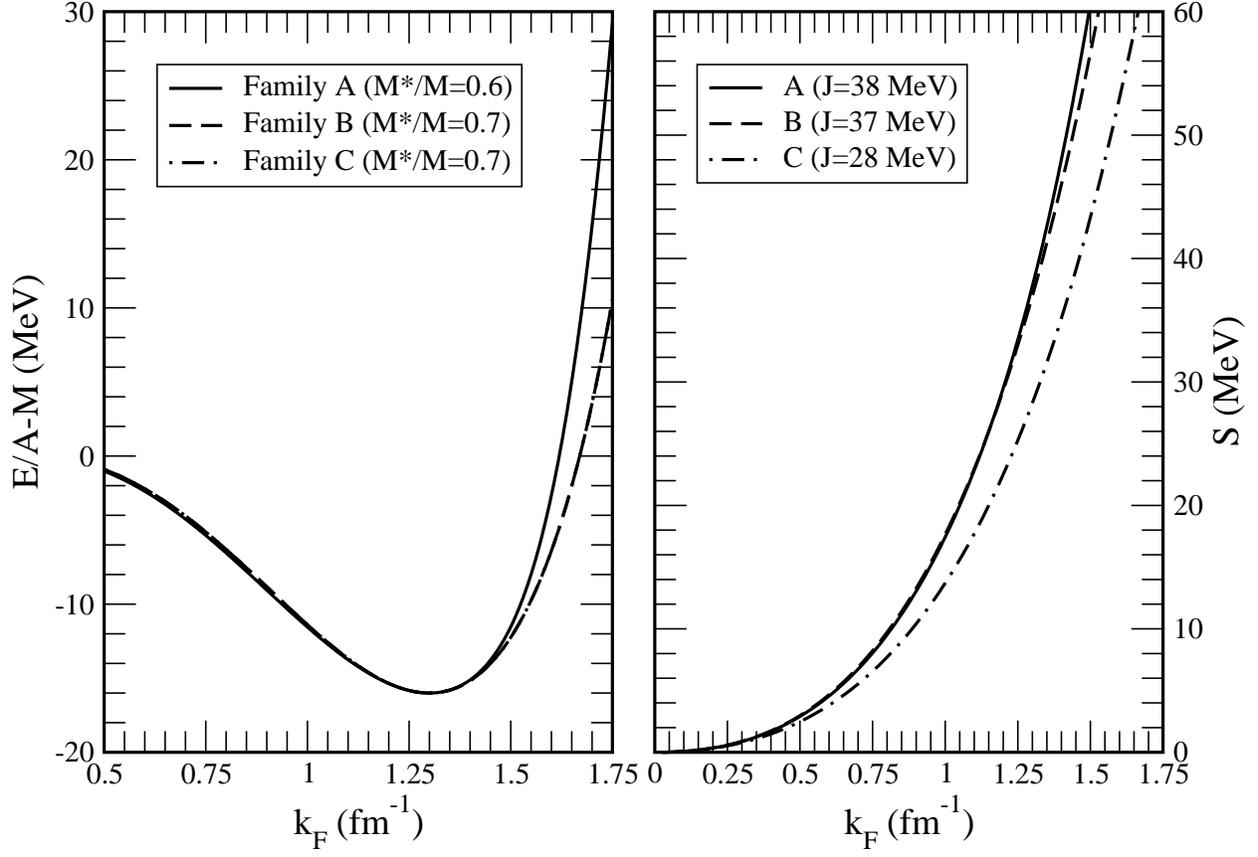}
 \caption{Equation of state for symmetric nuclear matter (left
	  panel) and the symmetry energy (right panel) as a
	  function of the Fermi momentum for the three families 
	  discussed in the text. In all the cases presented here 
	  the compression modulus was fixed at $K\!=\!250$~MeV.}
 \label{figure1}
\end{figure}

Within each family defined in Table~\ref{table1}, calculations of the
isoscalar monopole response have been performed using a compression
modulus in the physically acceptable range of $K\!=\!200\!-\!300$~MeV.
To illustrate the similarities and differences between these three
families, the equation of state for symmetric nuclear matter (left
panel) and the symmetry energy (right panel) are displayed in
Fig.~\ref{figure1} at $K\!=\!250$~MeV.  Clearly, the properties of
symmetric nuclear matter at saturation density are identical in all
three models. Further, having fixed the value of the effective nucleon
mass in symmetric nuclear matter, the full density dependence of the
symmetry energy is determined by one sole number: its value at
$k_F\!=\!1.15$~fm$^{-1}$.

\begin{figure}
 \includegraphics[width=0.9\linewidth]{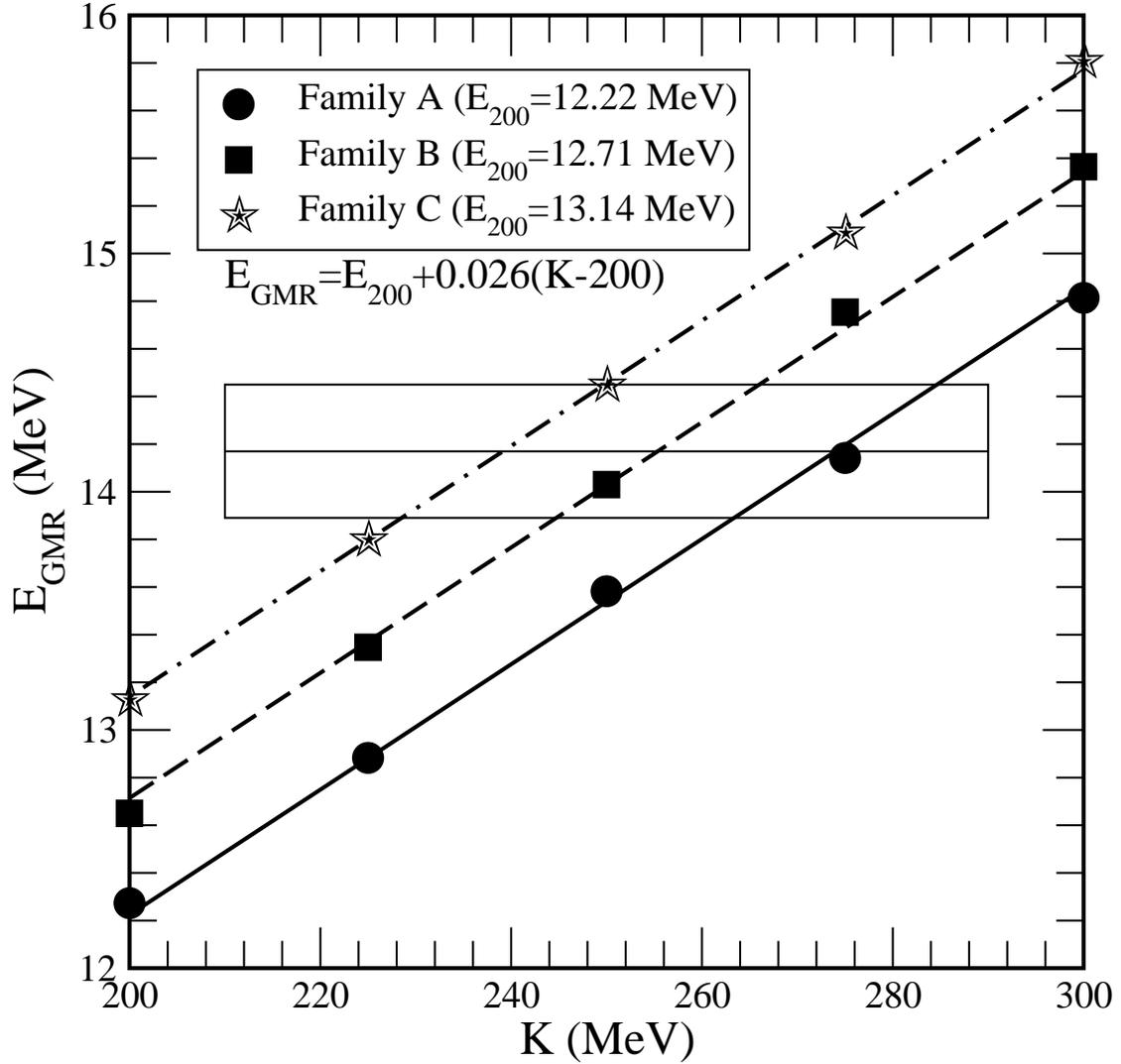}
 \caption{Energy of the isoscalar giant-monopole resonance
          as a function of the nuclear matter compression modulus
	  for the three families discussed in the text. The box 
	  displays the experimentally allowed range of 
	  $E_{\rm  GMR}\!=\!14.17\!\pm\!0.28$~MeV
	  \protect{\cite{Yo99}}.}
 \label{figure2}
\end{figure}

Results for the peak energy of the giant-monopole-resonance in
${}^{208}$Pb as a function of the nuclear incompressibility are 
listed in Table~\ref{table2} and displayed in Fig~\ref{figure2}.  
All calculations were performed using the nonspectral, relativistic
random-phase-approximation (RPA) approach of Ref.~\cite{Pi00}.  For
each family, there is a clear correlation between the compression
modulus and the energy of the GMR. Indeed, all of the results are well
represented (in this limited range of $K$) by a linear relation with a
``universal'' slope:
\begin{equation}
  E_{\rm GMR} = E_{200} + 0.026(K-200) \;,
 \label{EGMRvsK}
\end{equation}
where $E_{\rm GMR}$, $E_{200}$, and $K$ are all given in MeV. The
intercept is non-universal and given by:
$E_{200}\!=\!12.22$~MeV, $E_{200}\!=\!12.71$~MeV, and
$E_{200}\!=\!13.14$~MeV, for families A, B, and C, respectively.  

A few comments are now in order. First, the value of the slope
($0.026$) is obviously small. This suggests that even without
theoretical uncertainties, it would not be possible to determine the
compression modulus from the ${}^{208}$Pb measurement alone to better
than $\Delta E_{\rm GMR}/0.026$~MeV ($\Delta E_{\rm GMR}$ is
the experimental uncertainty).  At present, the best determination of
the peak position of the GMR is $E_{\rm
GMR}\!=\!14.17\!\pm\!0.28$~MeV~\cite{Yo99}, thereby resulting in an
uncertainty in the compression modulus of about $20$~MeV. Second, and
more importantly, the journey from the GMR to the compression modulus
is plagued by uncertainties unrelated to the physics of symmetric
nuclear matter. To illustrate this point we invoke, although never use
in any of the calculations, a semi-empirical formula based on a
leptodermous expansion of the nuclear incompressibility:
\begin{equation}
 K(A,I) = K + K_{\rm surf}/A^{1/3}
            + K_{\rm sym}I^{2}
	    + K_{\rm Coul}Z^{2}/A^{4/3} 
	    + \ldots \;,
 \label{LeptoDermous}
\end{equation}
where $K_{\rm surf}$, $K_{\rm sym}$, and $K_{\rm Coul}$ are empirical
surface, symmetry, and Coulomb coefficients and $I\!=\!(N\!-\!Z)/A$ is 
the neutron-proton asymmetry. The sizable contribution from the surface
term to $K(A,I)$ has been discussed recently by Patra, Vi\~nas,
Centelles, and Del Estal~\cite{Pa02} in the context of a relativistic
Thomas-Fermi theory so we limit ourselves to only a few comments. A
surface dependence is modeled here through a change in the value of
the effective nucleon mass (surface properties are also sensitive to
the $\sigma$-meson mass but this value has been chosen to reproduce
the rms charge radius of ${}^{208}$Pb). As shown in
Table~\ref{table1}, family~A uses an effective nucleon mass of
$M^{*}/M\!=\!0.6$ while family~B uses $M^{*}/M\!=\!0.7$; all other
input observables are identical. A larger $M^{*}$ generates a slightly
compressed single-particle spectrum and a correspondingly smaller
spin-orbit splitting. Consequences of this change in $M^{*}$ result in
a larger intercept, as displayed in Fig.~\ref{figure2}. Thus,
compression moduli of approximately $K\!=\!275$~MeV (for family~A) and
$K\!=\!250$~MeV (for family~B) are required to reproduce the
experimental energy of the GMR. Further, if one incorporates the
experimental error into this analysis, one concludes that ``best-fit''
relativistic mean-field models are consistent with a compression
modulus in the range $K\!=\!245\!-\!285$~MeV.

We now turn to the central idea behind this work, namely, how our
incomplete knowledge of the symmetry energy impacts on the the
extraction of the compression modulus. Let us then start by
considering two identical models, but with vastly different 
symmetry energies, that predict a compression modulus of
$K\!=\!250$~MeV. Further, for simplicity we assume that these two
models have identical surface and Coulomb properties so only the 
first and third term in Eq.~(\ref{LeptoDermous}) are relevant to this
discussion. Both models attempt to reproduce the ``experimentally''
accessible quantity:
\begin{equation}
 K_{208}\equiv\lim_{A\rightarrow\infty} K(A,I\!=\!0.212) 
        = K + K_{\rm sym}(0.212)^{2} + \ldots\;,
 \label{K208}
\end{equation}
defined as the compressibility of infinite nuclear matter at a
neutron-proton asymmetry identical to that of ${}^{208}$Pb (see
Table~\ref{table2}). The first model, having a very stiff symmetry
energy (that is, $K_{\rm sym}$ large and negative) reduces $K(A,I)$
from its $I\!=\!0$ value of $250$~MeV all the way down to, let us say,
$200$~MeV at $I\!=\!0.212$. Comparing this prediction to the assumed
experimental value of $K_{208}\!=\!225$~MeV, it is concluded that
the compression modulus of symmetric nuclear matter must be increased
to $K\!\simeq\!275$~MeV. The second model predicts a very soft symmetry
energy. So unrealistically soft, let us assume, that it generates no
shift in going from $I\!=\!0$ to $I\!=\!0.212$ ({\it i.e.,} $K_{\rm
sym}\!=\!0$).  In this case, the compression modulus must then be
reduced to $K\!=\!225$~MeV to reproduce the experimentally determined
value. Thus two models, originally identical as far as symmetric
nuclear matter is concerned, disagree in their final values of the
compression modulus due to an incomplete knowledge of the symmetry
energy. While the situation depicted in Fig.~\ref{figure2} might not
be as extreme, it does follow the trends suggested by the above
discussion. Indeed, family~C, with the softest symmetry energy,
generates the largest intercept and consequently predicts the smallest
compression modulus of the three families.

In summary, the impact of the poorly known density dependence of the
symmetry energy on the extraction of the compression modulus of
nuclear matter from the energy of the giant-monopole resonance in
${}^{208}$Pb was addressed.  The nuclear matter equation of state and
the distribution of isoscalar monopole strength in ${}^{208}$Pb were
computed using three different families of relativistic models
constrained to reproduce a variety of ground-state observables. For
each family the compression modulus was allowed to vary within the
physically acceptable range of $K\!=\!200\!-\!300$~MeV. The first
family (A), with an effective nucleon mass fixed at $M^{*}/M\!=\!0.6$
is, at least for $K\!=\!275$~MeV, practically indistinguishable from
the successful NL3 model of Ref.~\cite{La97}. The second family (B)
differs from the first in that the effective nucleon mass is increased
to $M^{*}/M\!=\!0.7$, thereby generating a slightly compressed
single-particle spectrum but still a robust phenomenology.  Finally,
the third family (C) is obtained from the second one by artificially
softening the symmetry energy in a ``poor-man's'' attempt at
simulating nonrelativistic Skyrme models.  When the peak energy of the
GMR is plotted against the compression modulus, a linear relation with
a universal slope is obtained. In contrast, the intercept is family
dependent and it is largest for the model with the softest symmetry
energy. Demanding agreement with the experimental value for the peak
energy fixes the compression modulus at: $K\!=\!275, 255,$ and
$240$~MeV, for families A, B, and, C, respectively. It is therefore
suggested that the discrepancy between relativistic and
nonrelativistic models in the prediction of the compression modulus of
nuclear matter may, at least in part, be due to our incomplete
knowledge of the symmetry energy.  At present, this issue can not be
resolved. Yet the proposed Parity Radius Experiment (PREX) at the
Jefferson Laboratory should provide a unique constraint on the density
dependence of the symmetry energy through a measurement of the neutron
skin of $^{208}$Pb. Such a measurement could have far-reaching
implications: from the determination of a fundamental parameter of the
equation of state ($K$) to the structure of neutron stars~\cite{Ho01}.

\begin{acknowledgments}
\vspace{-0.15in}
The author is grateful to the ECT* in Trento for their support 
and hospitality during the initial phase of this research. It 
is a pleasure to thank Profs. M. Centelles and X. Vi\~nas for 
many enlightening conversations. This work was supported in part 
by the U.S. Department of Energy under Contract No.DE-FG05-92ER40750. 
\end{acknowledgments}


\begin{table}
\caption{Empirical bulk observables used in the
	 determination of the coupling constants
	 and the scalar mass. The symmetry energy $J$
	 has been fixed at $k_F\!=\!1.15$~fm$^{-1}$ 
	 but the quantities in parenthesis represent 
	 its value at saturation density.}
 \label{table1}
 \begin{ruledtabular}
 \begin{tabular}{ccccccc}
 Family & $k_{\rm F}^{0}$~(fm$^{-1}$)  
        & $\epsilon_{0}$~(MeV) & $M^{*}/M$
        & $K$~(MeV) & $J$~(MeV) 
        & $r_{\rm ch}$~(fm) \\
 \hline
  A & $1.30$ & $-16$ & $0.6$ & $200\!-\!300$ & 
      $26(38)$ & $5.50\!\pm\!0.01$ \\ 
  B & $1.30$ & $-16$ & $0.7$ & $200\!-\!300$ & 
      $26(37)$ & $5.50\!\pm\!0.01$ \\ 
  C & $1.30$ & $-16$ & $0.7$ & $200\!-\!300$ & 
      $20(28)$ & $5.50\!\pm\!0.01$ \\ 
 \end{tabular}
 \end{ruledtabular}
\end{table}
\begin{table}
\caption{The compression modulus of symmetric nuclear matter,
	 the slope of the symmetry energy at saturation
	 density, the compression modulus for asymmetric 
	 ($I=0.212$) nuclear matter, the neutron skin of 
	 ${}^{208}$Pb, and the energy of the GMR in
	 ${}^{208}$Pb for the three families discussed 
	 in the text.} 
 \label{table2}
 \begin{ruledtabular}
 \begin{tabular}{cccccc}
  Family & $K$~(MeV) & $L$~(MeV)
         & $K_{208}$~(MeV) 
         & $R_{n}\!-\!R_{p}$~(fm)
         & $E_{\rm GMR}$~(MeV) \\
  \hline
   A & 200 & 120 & 184 & 0.28 & 12.27 \\ 
     & 225 & 120 & 203 & 0.28 & 12.88 \\ 
     & 250 & 119 & 224 & 0.28 & 13.58 \\ 
     & 275 & 119 & 246 & 0.28 & 14.14 \\ 
     & 300 & 119 & 268 & 0.28 & 14.81 \\
  \hline
   B & 200 & 108 & 187 & 0.25 & 12.65 \\ 
     & 225 & 108 & 208 & 0.25 & 13.35 \\ 
     & 250 & 108 & 230 & 0.26 & 14.03 \\ 
     & 275 & 108 & 252 & 0.26 & 14.75 \\ 
     & 300 & 107 & 276 & 0.26 & 15.36 \\
  \hline
   C & 200 &  82 & 190 & 0.19 & 13.13 \\ 
     & 225 &  82 & 212 & 0.19 & 13.80 \\ 
     & 250 &  82 & 235 & 0.19 & 14.45 \\ 
     & 275 &  82 & 258 & 0.19 & 15.09 \\ 
     & 300 &  82 & 282 & 0.19 & 15.81 \\
 \end{tabular}
\end{ruledtabular}
\end{table}
\end{document}